\documentclass[preprint,showpacs,showkeys,amsmath,amssymb]{revtex4-1}

\usepackage{xcolor}
\usepackage{setspace}
\usepackage{graphicx}
\usepackage{epsfig}
\usepackage[latin1]{inputenc}
\usepackage[active]{srcltx}
\usepackage{dcolumn}
\usepackage{bm}
\usepackage{epstopdf}

\begin{document}


\preprint{IST 4.2015-Pinheiro}

\title[]{Parametric resonance and particle stochastic interactions with a periodic medium}

\author{Mario J. Pinheiro}

\address{Department of Physics, Instituto Superior T\'{e}cnico - IST, Universidade de Lisboa - UL, Av. Rovisco Pais, \& 1049-001
Lisboa, Portugal}
\email{mpinheiro@tecnico.ulisboa.pt}




\date{\today}

\begin{abstract}
A non-markovian stochastic model shows the emergence of structures in the medium, a self-organization characterized by a relationship between particle's energy, driven frequency $\omega$ and a frequency of interaction with the medium $\nu$. The interaction determines its mass and this fine tuning results in an effective force given by $F_L=\hbar \omega^2 n(\lambda)/ c$, similar to the interaction force between photons and atoms. Condition for the particle-medium resonance is determined, with relevance to detect dark matter axion-like particles and the parametric resonance as a pop-up mechanism to turn fields into particles.
\end{abstract}

\maketitle


The study of physical systems with non-markovian statistical
properties has provided a natural basis for the understanding of
the role played by memory effects in such different fields as
anomalous transport in turbulent plasmas ~\cite{Ballescu95};
Brownian motion of macroparticles in complex fluids
~\cite{Amiblard96}; in the vortex solid phase of twinned
YBa$_2$Cu$_3$O$_7$ single crystals~\cite{Bekeris00}; simulating
the stochastic character of the laser fields ~\cite{Kofman}; the
rate of escape of a particle over a one-dimensional potential
barrier~\cite{Bray1,Bray2}.

Within a classical approach of an atomic process, we show in this paper that, whenever a particle undergoes a repetitive process, like a jumping process in a surrounding medium, a new type of force is exerted on it, the Lorentz invariant force~\cite{Vigier_2000}. The space evolution of a massless particle through a medium incorporates the space-time structure (e.g., topological, fractal) and the nature of motion.

In this study we embrace the concept of an information-rich manifold as the most reasonable heuristic framework in regard the non-Markovian propagation of a singularity in a complex manifold. The simple model introduced here
consists of a particle moving in a straight line for which we make no assumption about its mass (e.g., it's an {\it ab initio}  massless particle), jumping from one
site to another in a non-randomly structured field but, in the meanwhile, interacting with it in a stochastic process, and keeping memory of its "history".



In a non-markovian model the prediction about the next link
($x_{n+1}$) is defined in terms of mutually dependent random
variables in the chain ($x_1$, $x_2$,...,$x_n$). Consider a
particle jumping from one site to another in Euclidean space - non-Markovian singularities in a complex manifold.
The jumping sites are
assumed to be equidistantly distributed along the axis. Now, add
to this jumping process an oscillatory motion due to interaction
with a medium and characterized by stochasticity. The frequency of
oscillation around an equilibrium position between two jumps is
denoted by $\nu$, is homogeneous and isotropic (the {\it Zitterbewegungen}) and $\beta$ is the probability that each
oscillation in the past has to trigger a new oscillation in the
present.

Our simple dynamical process is introduced in a formal way, by
relating it to the probability that one oscillation from the
$M=m_0+...+m_{q-1}$ which occurred in the past generates $m$
oscillations at the $qth$ step, $Q_m[q(t)]$. Since we assume
$\beta$ is constant, this is an infinite memory model, meaning
that an oscillation which has occurred long time ago produces the
same effect as an oscillation which has occurred in the near past.
Lets introduce the probability density, $Q_n(t)dt$, that the $nth$
oscillation takes place in the interval of time $(t,t+dt)$ at
q$th$ step. Then we have the following integral in time
\begin{equation}\label{Eq1}
Q_{n+1}[q(t)]=\int_0^{q(t)} Q_{n}[q(t')] p_0[(q(t) - q(t')] d
q(t'),
\end{equation}
where $p_0(t-t')$ is the probability per unit time that the
$(n+1)st$ oscillation takes place in the time interval $(t,t+dt)$
given that the $nth$ oscillation took place at t'. Since the
particle is not allowed to come back and forth, there is no
entanglement in Eq.~\ref{Eq1}. Due to the hidden interactions the
particle undergo with the medium, we treat the time of an
oscillation as a random variable following a Poisson distribution
\begin{equation}\label{Eq2}
p_0(t-t') = \left\{
   \begin{array}{l l}
     0 &, \quad \text{if $(t-t') < \tau$}\\
     \nu d t \exp[-\nu(t-t')]  &, \quad \text{otherwise}.
   \end{array} \right.
\end{equation}
Here, $\nu$ is the frequency of an oscillation and $\tau$ is the
"dead" time. Designing by $\chi_n(s)$ and $\pi_0(s)$ the Laplace
transforms of $Q_n(t)$ and $p_0(t)$, resp., the convolution
theorem gives
\begin{equation}\label{Eq3}
\chi_{n+1}(s) = \chi_n(s)\pi_0(s).
\end{equation}
From this expression we obtain the recursive relation
\begin{equation}\label{Eq4}
\chi_n(s)=[\pi_0(s)]^{n-1} \chi_1(s).
\end{equation}
The evaluation of the transforms $\pi_0(s)$ and $\chi_1(s)$ gives
immediately
\begin{equation}\label{Eq5}
\pi_0(s)=\frac{\nu \exp(-(\nu+s)\tau)}{\nu + s},
\end{equation}
and
\begin{equation}\label{Eq6}
\chi_1(s)=\frac{\nu}{\nu +s},
\end{equation}
leading us to
\begin{equation}\label{Eq7}
\chi_n(s)=\nu^n \frac{\exp(-(n-1)(\nu+s]\tau}{(\nu + s)^n}.
\end{equation}
The inverse transform calculated using the Laplace inverse
theorem, gives the probability for the occurrence of $n$
oscillations at time $t$:
\begin{equation}\label{Eq8}
Q_n(t)= \left\{ \begin{array}{ll} \nu
\frac{\{\nu[t-(n-1)\tau]\}^{n-1} \exp(-\nu t)}{(n-1)!} &
\mbox{,$t>(n-1)\tau$}\\
  0 & \mbox{,$t<(n-1)\tau$}.
\end{array} \right.
\end{equation}
To simplify, we shall put $\tau=0$ and the probability density
that the $nth$ oscillation takes place in the interval of time
$(t,t+dt)$ reads
\begin{equation}\label{Eq9}
Q_n(t)d t=\frac{\nu(\nu t)^{n-1}}{(n-1)!}\exp(-\nu t) d t.
\end{equation}
It follows the probability density of occurrence of q jumps at
time $t$ is given by
\begin{equation}\label{Eq10}
\Psi_q(t) d t=\sum_{n=1}^{\infty} \xi_q(n) Q_n(t) d t,
\end{equation}
or, in complete form,
\begin{equation}\label{Eq11}
\Psi_q(t) d t = \sum_{n=1}^{\infty} \xi_q(n) \frac{\nu(\nu
t)^{n-1}}{(n-1)!} \exp(-\nu t) d t.
\end{equation}
Here, $\xi_q (n)$ is the probability to occur $n$ oscillations at
$qth$ jump. To evaluate $\xi_q(n)$ we first define $g_M(m_q)$, the
probability that M previous oscillations generate $m_q$
oscillations at $qth$ step~\cite{vlad}. The Bose-Einstein
distribution is favored since many oscillations can pertain to the
same step:
\begin{equation}\label{BED}
g_M(m_q) = \frac{(M + m_q - 1)!}{m_q!(M-1)!} \beta^{m_q} (1-\beta)^{M}.
\end{equation}

Introducing the conditional probability
$\varphi_q(m_q|m_{q-1},...,m_0)$ that at $qth$ step there are
$m_q$ oscillations provided that at the previous steps
$m_{q-1},...,m_0$ oscillations have occurred, subject to the
normalization condition
\begin{equation}\label{norcond}
\sum_{m_q} \varphi_q(m_q|m_{q-1},...m,m_0)=1,
\end{equation}
then, it can be shown ~\cite{vlad} that
\begin{displaymath}
\xi_q (n) = \sum_m \xi_0(m) (1-\beta)^{qm} [1-(1-\beta)^q]^{n-m}
\end{displaymath}
\begin{equation}\label{Eq13}
\frac{(n-1)!}{(m-1)! (n-m)!}.
\end{equation}


\begin{figure}
  \centering
  \includegraphics[width=3.4 in]{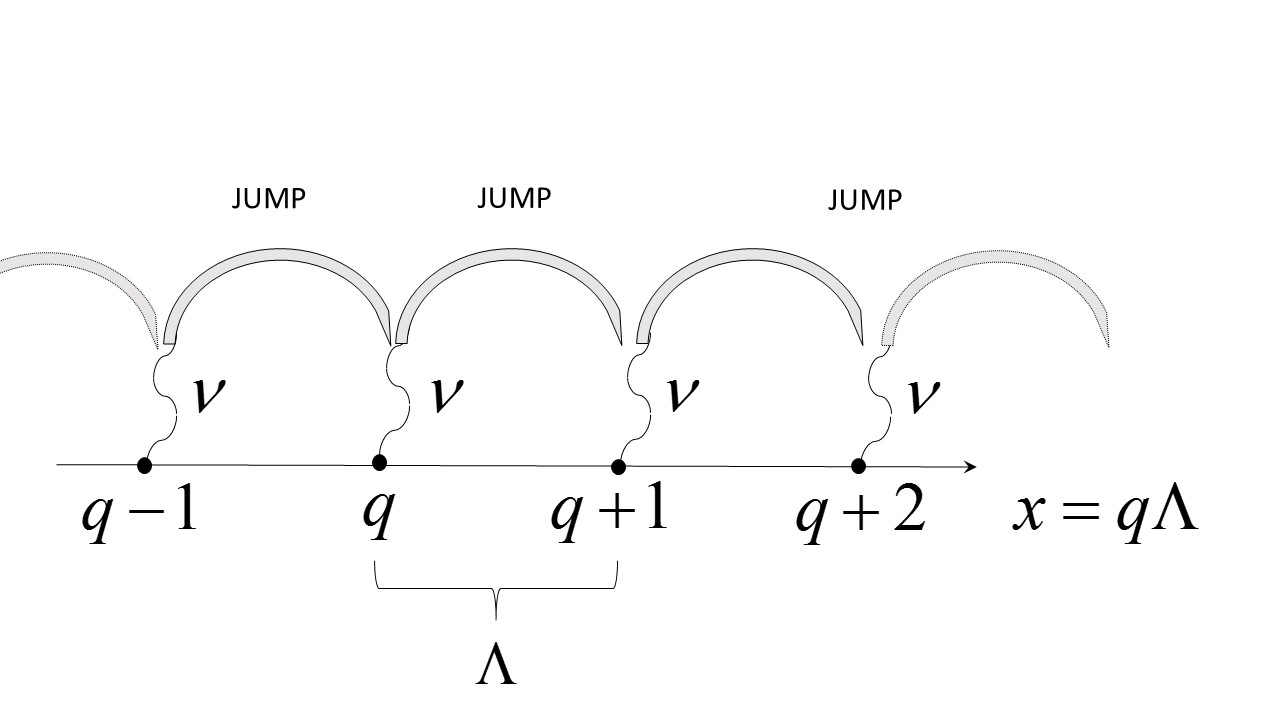}\\
  \caption{Particle in non-Markovian jumps in a lattice medium.}\label{Fig0}
\end{figure}

Therefore, the probability density of occurrence of q-jumps is
finally found to be
\begin{equation}\label{wp} \Psi_q(t) d t =
\frac{\alpha}{\nu} \exp(-\alpha t) \sum_m \xi_0(m) \frac{(\alpha
t)^{m-1}}{(m-1)!} d t,
\end{equation}
where we put $\alpha(q)\equiv (1-\beta)^q \nu$. It must be assumed
we know $\xi_0(m)$, that is the probability to occur $m$
oscillations from $t=0$ up to the first jump.

\begin{figure}
  \centering
  \includegraphics[width=3.4 in]{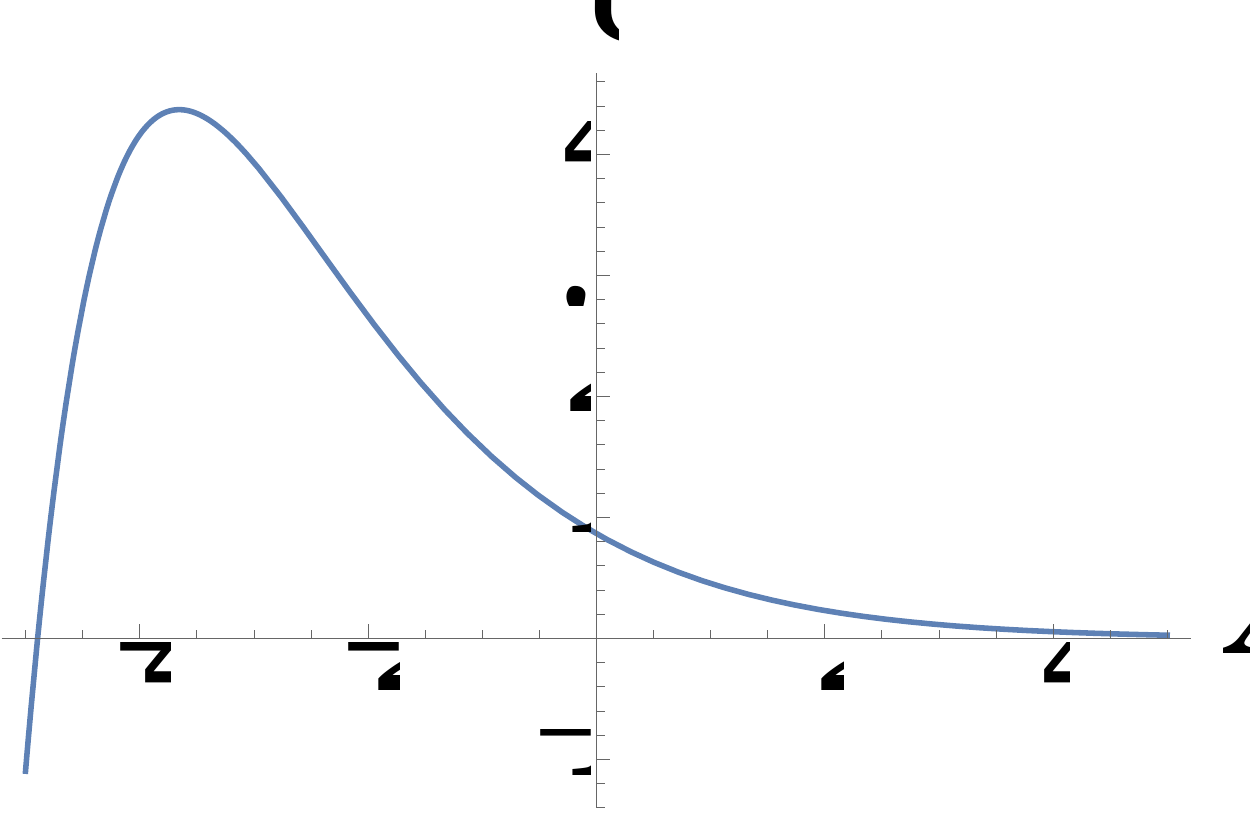}\\
  \caption{Wave-like soliton amplitude vs. position, in arbitrary units.}\label{Fig1}
\end{figure}

With the assumption of a Poisson distribution for $\xi_0(m)$, the
summation gives
\begin{equation}\label{gauss}
\sum_{m=0}^{\infty} \xi_0(m) \frac{(\alpha
t)^{m-1}}{(m-1)!}=\frac{1}{\sqrt{\lambda \alpha t}}
I_1(\sqrt{\lambda \alpha t}),
\end{equation}
where $I_1(x)$ is the first class modified Bessel function of
order 1. Hence, the final result for the probability of occurrence
of q-jumps between $t$ and $t+dt$ is given by
\begin{equation}\label{psi1}
\Psi_q(t) d t = \sqrt{\frac{\alpha}{\lambda \nu^2 t}} \exp(-\alpha
t) I_1(\sqrt{\lambda \alpha t}) dt.
\end{equation}
Eq.~\ref{psi1} is characterized by a temporal argument and, in
particular, for a sufficient number of steps, the limit
$\sqrt{\lambda x} \to 0$ is satisfied, and from the above we
obtain
\begin{equation}\label{prob2}
\Psi_q (t) d t \approx \frac{\alpha}{2} \exp[-\alpha t] d t.
\end{equation}
We have in view a deterministic particle system evolving according
to a local mapping in a space of equidistant sites. This
idealization lies in the Ehrenfest's equation describing the
quantum mechanical mean value of the particle position, and thus
avoids the solution of a much more complex problem~\cite{vlad}
which, in the problem here addressed, does not bring any further
substance. Hence, we can rewrite the above equation in the form
\begin{equation}\label{prob3}
\Psi(x,t) dt \approx \frac{\alpha}{2} \exp [-\alpha t] dt.
\end{equation}
According to the statistical interpretation of wave mechanics, the
probabilities are quadratic forms of a $\psi$ functions,
$\Psi_q(t)=|\psi_q(t)|^2$, with $\psi$ designating the associated
"wave". Therefore, we can seek a representation of the transport
process in terms of wave function. In fact, as we will see, this
is a far-reaching representation of the process. Fig.~\ref{Fig1} represents the wavefunction $\Psi=\frac{1}{\sqrt{t}}\exp(-t)*I_1(\sqrt{3t})$ of a soliton-like wave.

Inquiring for a convenient simplification of the complicated
initial function lead us to a simpler wave representation in which
a definite functional form as $x \pm vt$ is obtained. Reducing our
representation to harmonic waves in which way the energy
associated with the wave is expressed? Does the energy relation
$E=\hbar \omega$ and De Broglie relation hold on? Or does an
appropriated modification is at stake? By expanding the temporal
argument present in the exponential function in Eq.~\ref{prob3}
and retaining only terms of magnitude $\beta^2$ (higher order
terms are less important and it is harder to give them a physical
meaning), we obtain
\begin{equation}\label{dev}
(1- \beta)^q \nu t \approx \nu t  - \frac{\beta q \Lambda
\nu}{\Lambda} t + \frac{\hbar}{2} \frac{(\beta q \nu)^2}{\hbar
\nu} t + \mathcal{O}(\beta^3).
\end{equation}


The above expansion suggests the identification of some mechanical
properties of the particle, using the analogy with a
transversal wave in a vibrant string:
\begin{equation}\label{}
V \equiv \frac{\nu l}{2 \pi n'}, ~\mbox{with n'=1,2,3,...},
\end{equation}
assuming a non-dispersive medium. We denote by $l \equiv q \Lambda$ the distance travelled by the particle from a fixed point $O$ of the x-axis after time $t$ and $\nu$ is the number of
cycles per second loosed on a given space position, both
quantities as seen by an observer at rest in the lattice. The wave number is defined by
\begin{equation}\label{}
K \equiv\frac{\beta 2 \pi n}{\Lambda},
\end{equation}
where $\beta$ is the probability that each oscillation in the past has to trigger a
new oscillation in the present. We also obtain $\omega$:
\begin{equation}\label{}
\omega = \beta q \nu.
\end{equation}
Notice that when the number of jumps is $q=1$ and the probability is equal to the unity, $\beta=1$, then $\omega = \nu$, otherwise, they acquire different values.
Attributing physical meaning to the parameters permits to identify the third term on the right-hand side of
Eq.~\ref{dev} with the energy carried by the particle:
\begin{equation}\label{kla}
E = \frac{\hbar}{2 \nu} \left( \beta q \nu \right)^2.
\end{equation}


We have made so far no hypothesis about the mass of the ideal particle. But due to the jumping and interaction with the medium, using Einstein's relationship, $E=\overline{m}c^2$, the particle energy is equivalent to a mass, its own ``mass", and the energy content is given by
\begin{equation}\label{}
E = \frac{1}{2}\frac{(\beta q \nu)^2}{\nu} \hbar.
\end{equation}
Therefore,
\begin{equation}\label{Broglie2}
\overline{m} = \frac{\hbar \omega^2}{2\nu c^2}.
\end{equation}
The above equation represents a structural relationship between energy, driven frequency $\omega$ and characteristic frequency of interaction with the medium, $\nu$ and we may notice that the non-markovian character of the stochastic process is more intrinsically related to the nature of the medium rather than the past history of the particle.

As shown before, for consistence $\omega \equiv \beta q \nu$, and then
from Eq.~\ref{Broglie2} we have
\begin{equation}\label{Eq14}
E=\frac{\hbar \omega^2}{c}\frac{c}{2\nu}=\left( \frac{\hbar \omega^2}{c} \right)\lambda_0.
\end{equation}
Notice that the factor $2$ appears since the particle has two degrees of freedom for the transversal vibration (in a three dimensional space), see also the discussion in Ref.~\cite{Osche_2011}. Hence, the frequency $\nu_0=2 \nu$ can be associated to the {\it Zitterbewegung}, recently experimentally observed~\cite{Roos_2010}, and therefore $c=\nu_0 \lambda_0$, with $\lambda_0=\lambda n$. In this case, we obtain
\begin{equation}\label{Eq14a}
E=\frac{\hbar \omega^2}{\nu_0}.
\end{equation}


The presence of two distinct frequencies in Eq.~\ref{Eq14a} suggests the possible occurrence of parametric resonance effect between the particle and the medium. Indeed, experiments have shown that particles possess an internal clock (the well-known hypothesis advanced by De Broglie) characterized by $\omega$ and when they interact with the medium, characterized by $\nu_0$, possible resonance may occur. This effect was shown in a channeling experiment with $\sim 80$ MeV electrons traversing a 1-$\mu$m thick silicon crystal aligned with the $<110>$ direction (nuclear scattering effects are stronger than in random direction). When the frequency of atomic collisions matches the internal clock frequency, the rate of electron transmission shows an 8 $\%$ dip within $0.5\%$ of the resonance energy~\cite{Roussel_2005}.
This idea could serve as a critical test bed for particle physics phenomena that seems to share common points with the parametric resonance effect between particles and the space-time lattice to detect dark matter axion-like particles~\cite{Gamboa_2015} or a periodic medium like graphene to investigate the origin of half-spin quarks~\cite{Regan_2012}, or even parametric resonance as a pop-up mechanism to turn fields into particles~\cite{Shaw_1999}, just to cite a few.


The interaction of the singularity particle with the medium develops a resistance (inertia) and gives rise to a new type of force shown in Eq.~\ref{Eq14}, also obtained with a different approach in Ref.~\cite{Vigier_2000} (and named by J. P. Vigier, the Lorentz invariant force):
\begin{equation}\label{Eq15}
F_L=\frac{\hbar \omega^2}{c}n(\lambda),
\end{equation}
for a medium with refraction index $n(\lambda)$. This force (or energy) actuating on a particle is at the origin of the mass of the particle. The analogy with a vibrating string allows the conjecture of the existence of higher harmonics.

The medium perturbation is characterized by $\nu$, a particular property of the particle surrounding medium from where it emerges the inertia of matter by means of the coefficient $\overline{m}$ and introducing a nonlinearity that produces a different pattern that the one
conceptualized by the quantum mechanical expression for a photon packet
($E=\hbar \omega$). Note that Eq.~\ref{Broglie2} is consistent
with the De Broglie relation for free particles (planar waves),
since then $\omega=\nu_0$. Otherwise, when the
interaction with the surrounding medium imposes a nonlinear dynamics, a new relationship is set-up, Eq.~\ref{Broglie2}. This scheme leads us to a description of quanta as embedded within a complex manifold, reminding the appearance of discrete objects as part of the medium, much like propagating soliton-like waves in a fluctuating, information-rich energy field. De Broglie~\cite{Broglie} and David Bohm~\cite{Bohm} were proponents of the Guided Wave Theory which proposes that particle's mass, as also it appears in our analysis, is not an intrinsic property of matter but an outcome of its interaction with a periodic medium. This simple and apparently universal mechanism is considered in contemporary cosmology, in the initial reheating process after inflation, when an explosive particle production takes place due to induce parametric resonance~\cite{Linde_2000}, and may explain pion production in a nonequilibrium chiral phase transition~\cite{Minakata_2001}.

\begin{acknowledgments}
The author gratefully acknowledge partial financial support by the FCT under contract ref. SFRH/BSAB/1420/2014.
\end{acknowledgments}


\end{document}